%% file: jrn_vmf_doppler.tex
\documentclass[journal,comsoc]{IEEEtran}
\usepackage[T1]{fontenc}

\usepackage[pdftex]{graphicx}
\graphicspath{{./figs/}}

\usepackage[acronym]{glossaries}
\loadglsentries[acronym]{acronyms}

\usepackage{amsmath, amssymb}
\providecommand{\abs}[1]{\lvert#1\rvert} 
\providecommand{\norm}[1]{\left\lVert#1\right\rVert} 

\usepackage{color}

\ifCLASSOPTIONcompsoc
 \usepackage[caption=false,font=normalsize,labelfont=sf,textfont=sf]{subfig}
\else
 \usepackage[caption=false,font=footnotesize]{subfig}
\fi

\newcommand\copyrighttext{%
  \textcopyright \the\year{} IEEE.
  Personal use of this material is permitted. \\
  Permission from IEEE must be obtained for all other uses, including reprinting/republishing this material for advertising or promotional purposes, collecting new collected works for resale or redistribution to servers or lists, or reuse of any copyrighted component of this work in other works.
}

\begin{document}
\title{Doppler Power Spectrum in Channels with \\von Mises-Fisher Distribution of Scatterers}

\author{Kenan~Turbic,~\IEEEmembership{Member,~IEEE,}
  Martin~Kasparick,
  and S\l{}awomir~Sta\'{n}czak,~\IEEEmembership{Senior Member,~IEEE}%
\thanks{%
%
This work was partially supported by the Federal Ministry of Education and Research (BMBF, Germany) in the “Souverän. Digital. Vernetzt.” programme, joint Project 6G-RIC; project identification numbers: 16KISK020K and 16KISK030.
It was developed within the scope of COST Action CA20120 (INTERACT).
%
\textit{(Corresponding author: K. Turbic.)}
}%
\thanks{The authors are with the Wireless Communications and Networks Department, Fraunhofer Institute for Telecommunications, Heinrich Hertz Institute (HHI), Berlin, 10587 Germany (e-mail: kenan.turbic@hhi.fraunhofer.de, martin.kasparick@hhi.fraunhofer.de, slawomir.stanczak@hhi.fraunhofer.de).}%
\thanks{S. Stanczak is also with Technische Universit{\"a}t Berlin, Berlin, Germany.}%
}


\IEEEpubid{%
  \begin{minipage}{0.8\textwidth}\ \\[10mm] \centering
      \footnotesize \copyrighttext
  \end{minipage}
}

\maketitle

\begin{abstract}
This letter presents an analytical analysis of the Doppler spectrum in \gls{vmf} scattering channels.
A simple closed-form expression for the Doppler spectrum is derived and used to investigate the impact of the \gls{vmf} scattering parameters, i.e., the mean direction and the degree of concentration of scatterers.
The spectrum is observed to exhibit exponential behavior for mobile antenna motion parallel to the mean direction of scatterers, while conforming to a Gaussian-like shape for the perpendicular motion.
The validity of the obtained results is verified by comparison against the results of Monte Carlo simulations, where an exact match is observed.
\end{abstract}

\medskip 
\begin{IEEEkeywords}
  Wireless communications, Mobile channel, Statistical model, von Mises-Fisher distribution, Doppler spectrum
\end{IEEEkeywords}

\glsresetall

%
\IEEEpeerreviewmaketitle

\section{Introduction}
\label{sec:intro}

\IEEEPARstart{D}{ue} to its flexibility, good fit to experimental data and the ability to approximate arbitrary 3D scattering by employing a model mixture \cite{Mammasis2009}, \gls{vmf} distribution is a widely adopted scattering model in the literature \cite{Zhu2018a, Bian2019}.
It provides distinct advantages over the traditional 2D scattering models \cite{Clarke1968, Abdi2002}, neglecting the vertical aspect of propagation, and their popular 3D extensions \cite{Aulin1979, Parsons1991}, confining the main scattering direction to the horizontal plane.
This makes the \gls{vmf} model a preferable choice for the analysis of multi-antenna communication systems \cite{Wang2023, Pizzo2022a}.

Despite the model's wide adoption, the analysis of the second-order channel characteristics under the \gls{vmf} scattering in the literature has been mostly restricted to numerical integration approaches.
While important analytical results for the spatial and temporal correlation functions \cite{Turbic2024, Zeng2024} and for the level-crossing rate \cite{Zeng2024a} were reported recently, such results are not yet available for the Doppler power spectrum.

To address this gap, this letter investigates the Doppler power spectrum in mobile channels with \gls{vmf} scattering.
As the main contribution of this work, we derive a novel closed-form solution for the Doppler power spectrum directly from the scattering distribution.
The obtained result is used to investigate the impact of the scattering parameters, i.e., the mean direction and the degree of concentration of scattering.
The simplicity of the derived analytical solution contrasts the spherical waveform expansion framework result \cite{Ho2005}, applicable to \gls{vmf} scattering (albeit results were not reported), involving expressions with infinite sums of special functions and offering little advantage over the numerical integration approach. \looseness=-1

The rest of this letter is structured as follows.
Section \ref{sec:ch_mod} presents the adopted channel model with the underlying assumptions, while Section \ref{sec:dopp_pdf_deriv} presents derivation of the corresponding Doppler spectrum.
The impact of the \gls{vmf} distribution parameters on the Doppler spectrum is analyzed in Section \ref{sec:results}, and the paper is concluded in Section \ref{sec:conclusions}.

\section{Channel Model}
\label{sec:ch_mod}
By considering multipath propagation between the \gls{tx} and the \gls{rx}, the complex channel transmission coefficient can be written as \cite{Molisch2011book}
\begin{align}
  H(t, f)
  &= \sum_{n=1}^{N} A_n e^{-j2\pi f \tau_0^n} e^{-j 2\pi f_D^n t }
  \label{eq:ch_coeff}
\end{align}
where
\begin{description}
	\item[$t$] time;
	\item[$f$] frequency;
	\item[$N$] number of multipath components;
	\item[$A_n$] their amplitudes;
	\item[$\tau_0^n$] initial propagation delays;
	\item[$f_D^n$] Doppler frequency shifts.
\end{description}

The scattering is assumed to occur in the far field of the mobile antenna with the multipath components' amplitudes, \glspl{doa} and delays assumed to be approximately constant over local areas, i.e., several tens or hundredsof wavelengths in size \cite{Molisch2011book}.
Together with linear motion of the mobile antenna, i.e., with constant speed and direction, these assumptions yield a stationary channel model \cite{Patzold2012book}. \looseness=-1

The initial propagation delays, associated with the apparently random propagation path lengths observed at the start of the channel observation period ($t=0$), result in random phase shifts of multipath components arriving at the \gls{rx}.
For narrowband communications, these initial phases are typically modeled as uniform random variables, i.e., $\varphi_0^n \sim \mathcal{U}(0, 2\pi)$.

\IEEEpubidadjcol

The time-variant phase terms account for the Doppler effect, associated with changes in the propagation path lengths due to mobile antenna mobility.
For constant speed motion, these changes are linear with rates corresponding to the Doppler frequency shifts exhibited by multipath components \cite{Molisch2011book}, i.e.
\begin{align}
	f_D^n
	&= \frac{1}{\lambda} \, \hat{\mathbf{k}}_n^T \mathbf{v}
	\label{eq:dopp_freq}
\end{align}
where
\begin{description}
  \item[$\lambda$] wavelength;
  \item[$\mathbf{v}$] mobile antenna velocity vector;
  \item[$\hat{\mathbf{k}}_n$] \gls{doa}%
  \footnote{In this work we arbitrarily chose to consider a static \gls{tx} and a mobile \gls{rx}. However, the same results apply if the roles are exchanged.}
  unit vector, i.e.
  \begin{align}
    \hat{\mathbf{k}}_n
    = (\cos\phi_n\cos\psi_n, \sin\phi_n\cos\psi_n, \sin\psi_n)^T
    \label{eq:doa}
  \end{align}
  \item[$\phi_n$] \gls{aaoa};
  \item[$\psi_n$] \gls{eaoa};
  \item[$(\,.\,)^T$] vector transpose operation.
\end{description}

The random \glspl{doa} observed at an arbitrary \gls{rx} position are represented by the \gls{vmf} distribution, with the \gls{pdf} given by \cite{Mardia2000book}
\begin{align}
	p_{\phi\psi}(\phi, \psi)
  &=
  \frac{\kappa \cos\psi}{4\pi\sinh{\kappa}}
  e^{\kappa\left[ \cos\mu_\psi \cos\psi \cos(\phi - \mu_\phi) + \sin\mu_\psi \sin\psi\right]}
  , \notag \\[1mm]
  &\phantom{=}\abs{\phi} \leq \pi, \;
  \abs{\psi} \leq \pi/2
	\label{eq:vmf_pdf}
\end{align}
where
\begin{description}
	\item[$\mu_\phi$] mean \gls{aaoa};
	\item[$\mu_\psi$] mean \gls{eaoa};
	\item[$\kappa$] spread parameter.
\end{description}
The parameters $\mu_\phi$ and $\mu_\psi$ specify the mean direction of scattering and $\kappa$ specifies the degree of angular concentration.
For $\kappa=0$, the \gls{vmf} distribution becomes uniform over the sphere and the scattering is isotropic, while for $\kappa\rightarrow\infty$ the scattering becomes deterministic, collapsing to a single scatterer in the direction given by ($\mu_\phi$, $\mu_\psi$).

For a large number of multipath components the channel coefficient in \eqref{eq:ch_coeff} exhibits a complex Gaussian distribution, as follows according to the central limit theorem.
Assuming no dominant multipath components are present, the \gls{rx} signal undergoes Rayleigh fading.
This case is considered in herein.

It is important to point out that, for a single-frequency harmonic signal emitted by the \gls{tx} antenna, the \gls{rx} signal consists of multiple frequency-shifted components.
The exhibited frequency shifts are given by \eqref{eq:dopp_freq} and depend on the underlying scattering distribution.
The resulting energy distribution across frequency is characterized by the Doppler \gls{psd}, derived in the following section.

\section{Doppler Power Spectrum}
\label{sec:dopp_pdf_deriv}
When normalized to unit power, the Doppler \gls{psd} corresponds to the Doppler frequency \gls{pdf} imposed by the underlying scattering distribution \cite{Molisch2011book}.
To derive the relationship between the scattering distribution and the Doppler \gls{pdf}, we furst have to establish the important geometrical relationships.

\begin{figure}[!t]
	\centering
	\includegraphics[width=8cm]{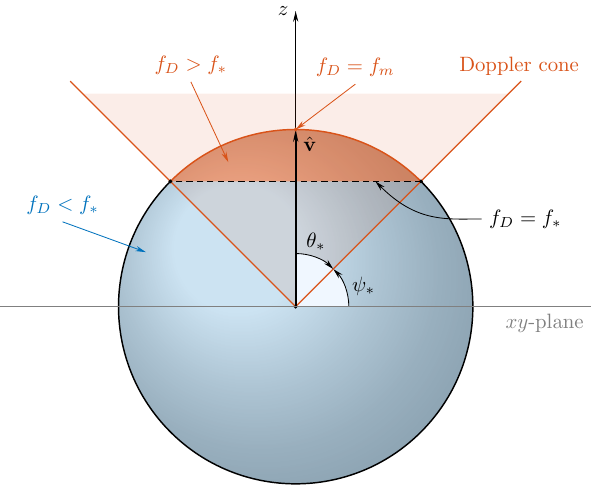}
	\caption{Doppler cone associated with Doppler shift frequency $f_{*}$, for a coordinate system chosen such that the $z$-axis is aligned with the mobile antenna velocity vector.}
	\label{fig:doppler_cone}
\end{figure}

To this end, we recall that all multipath components arriving at the \gls{rx} with the same inclination angle to the mobile antenna velocity vector exhibit the same Doppler frequency shift.
These directions define a circular cone with its apex at the mobile antenna position and the axis parallel to the velocity vector \cite{Narasimhan1999}.
Fig.~\ref{fig:doppler_cone} illustrates the Doppler cone for a Doppler frequency $f_{*}$, with the coordinate system chosen such that the $z$-axis is aligned with the mobile antenna velocity vector.%
\footnote{This choice of the coordinate system results in no loss of generality, since the relative geometry is preserved.}

The figure also shows a unit sphere, over which the distribution of \glspl{doa} is defined.
The points on the circular intersection of this sphere and the Doppler cone, indicated by the black dashed line, correspond to the unit \gls{doa} vectors associated with the same Doppler frequency shift, i.e.
\begin{align}
  f_{*}
  &= f_m \cos\theta_{*}
  \label{eq:doppler_beta}
\end{align}
where
\begin{description}
  \item[$f_m$] maximum Doppler shift, i.e., $f_m = \norm{\mathbf{v}}/\lambda$;
  \item[$\theta_{*}$] angle between the \gls{doa} and the motion direction.
\end{description}
The area of the sphere within the Doppler cone (shown in red) corresponds to the directions for which the Doppler frequencies are higher than $f_{*}$, and the sphere area outside of the cone (shown in blue) corresponds to the lower Doppler frequencies. \looseness=-1

Based on these observations, the \gls{cdf} of the Doppler frequency can be expressed as
\begin{align}
  F_{f_D}\left(f_{*}\right)
  = \int_{-\pi/2}^{\psi_{*}} \int_{-\pi}^{\pi} p_{\phi\psi}(\phi, \psi) d\phi d\psi
  \label{eq:dopp_cdf0}
\end{align}
where
\begin{description}
  \item[$\psi_{*}$] angle complementary to $\theta_{*}$, i.e., $\psi_{*} = \frac{\pi}{2}-\theta_{*}$ (Fig.~\ref{fig:doppler_cone}).
\end{description}
By inserting \eqref{eq:vmf_pdf} into \eqref{eq:dopp_cdf0} and applying \cite[Eq.~3.338.4]{Gradshteyn2007book} to solve the inner integral in the azimuth angle $\phi$, we obtain
\begin{align}
	F_{f_D}\left(f_{*}\right)
  &= \frac{\kappa}{2\sinh{\kappa}}
	\int_{-\pi/2}^{\psi_{*}} I_0\left(a \cos\psi  \right) e^{b \sin\psi} \cos\psi d\psi
  \label{eq:dopp_cdf1}
\end{align}
where
\begin{align}
	a &= \kappa \, \sqrt{\hat{k}_x^2+\hat{k}_y^2}
  \label{eq:B_var0}
  \\
  b &= \kappa \, \hat{k}_z
  \label{eq:c_var0}
\end{align}
with $\hat{k}_{x/y/z}$ denoting the components of the mean \gls{doa} vector,
\begin{align}
  \mathbf{k}_\mu = (\cos\mu_\phi\cos\mu_\psi, \sin\mu_\phi\cos\mu_\psi, \sin\mu_\psi)^T
  \label{eq:doa_mean}
\end{align}

The Doppler \gls{pdf} can be obtained as the derivative of \eqref{eq:dopp_cdf1} with respect to $f_{*}$.
Applying the Leibnitz integral rule yields%
\footnote{Only the upper limit of integration in \eqref{eq:dopp_cdf1} is a function of $f_{*}$, given by \eqref{eq:psi_beta}, while the integrand and the lower limit are invariant with respect to $f_{*}$.}
\begin{align}
  p_{f_D}(f_{*})
  &= \frac{\kappa}{2\sinh{\kappa}} I_0\left(a \cos\psi_{*}  \right) e^{b\sin\psi_{*}} \cos\psi_{*} \frac{d\psi_{*}}{df_{*}}
  \label{eq:dopp_pdf0}
\end{align}
From \eqref{eq:doppler_beta}, $\psi_{*}$ can be expressed as
\begin{align}
  \psi_{*} = \arcsin\left( \frac{f_{*}}{f_m} \right) , \qquad \abs{f} \leq f_m
  \label{eq:psi_beta}
\end{align}
and we obtain
\begin{align}
  \frac{d\psi_{*}}{df_{*}} = \frac{1}{f_m \sqrt{1-\left( \frac{f_{*}}{f_m} \right)^2}}
  \label{eq:psi_beta_deriv}
\end{align}

Replacing \eqref{eq:psi_beta_deriv} in \eqref{eq:dopp_pdf0}, while recalling the complementarity of $\psi_{*}$ and $\theta_{*}$ and using \eqref{eq:doppler_beta} to express $\cos\psi_{*}$ and $\sin\psi_{*}$ in terms of $f_{*}$, yields%
\footnote{Hereafter, we use $f$ instead of $f_{*}$ to simplify the notation.}
 ($\abs{f} \leq f_m$)
\begin{align}
  p_{f_D}(f)
  &= \frac{1}{2 f_m}\frac{\kappa}{\sinh{\kappa}} I_0\left(a \sqrt{1-\left( \frac{f}{f_m} \right)^2}  \right) e^{b\frac{f}{f_m}}
  \label{eq:dopp_pdf1}
\end{align}
where
\begin{description}
  \item[$I_0(\,.\,)$] modified Bessel function (first kind, zeroth order).
\end{description}
By observing from \eqref{eq:B_var0} and \eqref{eq:c_var0} that $a$ and $b$ solely depend on the components of the mean \gls{doa} vector perpendicular and parallel to the motion direction, respectively, we can write:
\begin{align}
  a &= \kappa \sqrt{1 - (\hat{\mathbf{k}}_\mu^T \hat{\mathbf{v}})^2}
  \label{eq:B_var}
  \\
  b &= \kappa \; (\hat{\mathbf{k}}_\mu^T \hat{\mathbf{v}})
  \label{eq:c_var}
\end{align}
where
\begin{description}
  \item[$\mathbf{\hat{v}}$] motion direction unit vector, i.e. $\mathbf{\hat{v}}=\mathbf{v}/\norm{\mathbf{v}}$.
\end{description}
Inserting these in \eqref{eq:dopp_pdf1} finally yields the general expression for the Doppler \gls{pdf} in \gls{vmf} scattering channels, i.e.
\begin{align}
  p_{f_D}(f)
  &=
  \frac{1}{2f_m} \frac{\kappa}{\sinh \kappa}
  e^{\kappa \, (\mathbf{k}_\mu^T \mathbf{\hat{v}}) \frac{f}{f_m} } \; \times
  \notag \\
  &\phantom{=}
  I_0\left( \kappa\sqrt{1 - (\hat{\mathbf{k}}_\mu^T \mathbf{\hat{v}})^2} \sqrt{1 - \left(\frac{f}{f_m}\right)^2} \right),
  \quad \abs{f}\leq f_m
  \label{eq:dopp_pdf}
\end{align}
%
In contrast to the results for the 2D scattering \cite{Clarke1968, Abdi2002}, the Doppler \gls{pdf} in \eqref{eq:dopp_pdf} is finite for $f = \pm f_m$, with
\begin{align}
  p_{f_D}(\pm f_m)
  &=
  \frac{1}{2f_m} \frac{\kappa}{\sinh \kappa}
  e^{\pm \kappa \, (\mathbf{k}_\mu^T \mathbf{\hat{v}})}
  \label{eq:dopp_pdf_max}
\end{align}

For the special case of isotropic scattering ($\kappa=0$), the Doppler \gls{pdf} is uniform, i.e.
\begin{align}
  p_{f_D}(f)
  &=
  \frac{1}{2f_m},
  \qquad \abs{f}\leq f_m
\end{align}
On the other hand, for $\kappa\rightarrow\infty$ the Doppler \gls{pdf} becomes
\begin{align}
  p_{f_D}(f) = \delta( f-f_{\mu} )
\end{align}
where
\begin{description}
  \item[$f_{\mu}$] Doppler shift for the mean \gls{doa} direction, i.e.
  \begin{align}
    f_{\mu}
    = f_m (\mathbf{k}_\mu^T \mathbf{\hat{v}})
    = f_m \cos\beta
  \end{align}
  \item[$\beta$] angle between the vectors $\mathbf{k}_\mu$ and $\mathbf{\hat{v}}$.
\end{description}
Therefore, scattering is deterministic in this case, with a single scatterer in the direction corresponding to the mean \gls{doa}.

\section{Results analysis}
\label{sec:results}
In this section we employ the obtained analytical results to investigate the impact of the \gls{vmf} scattering parameters.
To this end, the effect of the mean \gls{doa} direction and the concentration of scatterers around this direction is analyzed.

\begin{figure}[!t]
	\centering
  \includegraphics[scale=1]{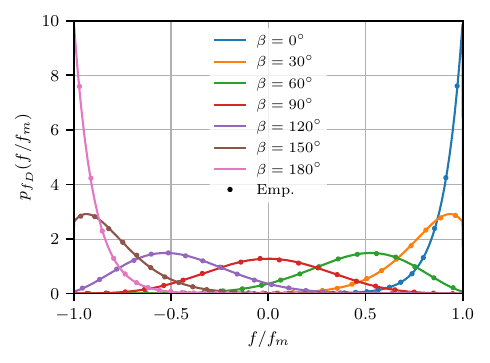}
	\caption{Doppler \gls{pdf} as a function of the angle $\beta$ between the velocity vector and mean \gls{doa} ($\kappa=10$); dots represent results obtained via simulation.}
	\label{fig:pdf_vs_beta}
\end{figure}

Fig.~\ref{fig:pdf_vs_beta} shows the Doppler \gls{pdf}, i.e., Doppler spectrum normalized to unit power, for different relative angles between the mobile antenna motion direction and the mean \gls{doa}.
For motion perpendicular to the mean \gls{doa} ($\beta=90^\circ$), the Doppler \gls{pdf} is observed to be an even function.
As the motion direction inclines towards the mean \gls{doa} ($\beta<90^\circ$) or opposite to it ($\beta>90^\circ$), the \gls{pdf} becomes skewed towards the maximum or minimum Doppler frequency, respectively.
For motion parallel to the the mean \gls{doa} ($\beta=0^\circ, 180^\circ$), the Doppler \gls{pdf} exhibits an exponential behavior.
This follows as the Bessel function term in \eqref{eq:dopp_pdf} is equal to one in this case, and the exponential term governs the shape of the Doppler \gls{pdf}. \looseness=-1

\begin{figure}[!t]
	\centering
	\subfloat[$\beta=0^\circ$.]{
		\includegraphics[scale=1]{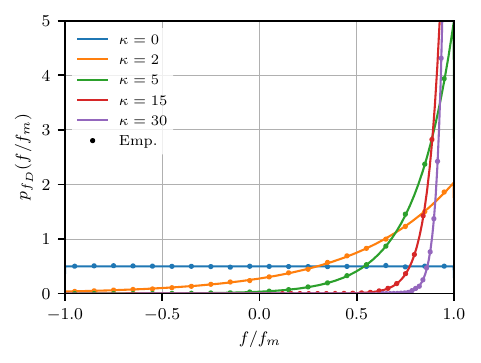}
		\label{fig:pdf_vs_kappa_beta0}
  }
	\hfil
	\subfloat[$\beta=90^\circ$.]{
		\includegraphics[scale=1]{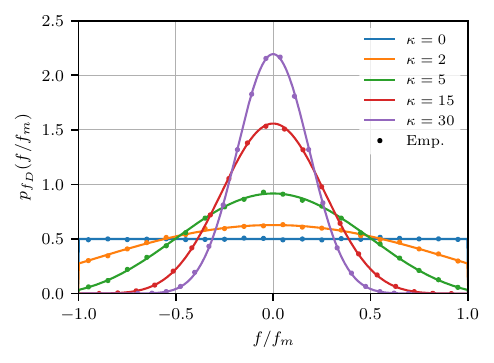}
		\label{fig:pdf_vs_kappa_beta90}
	}
	\caption{Doppler \gls{pdf} as a function of the spread parameter $\kappa$ for $\beta=90^\circ$ and $\beta=0^\circ$; dots represent results obtained via simulation.}
	\label{fig:pdf_vs_kappa}
\end{figure}

Fig.~\ref{fig:pdf_vs_kappa} shows the Doppler \gls{pdf} for different values of the angular spread parameter $\kappa$, in the case when the mean \gls{doa} is parallel (Fig.~\ref{fig:pdf_vs_kappa_beta0}) or perpendicular to the motion direction (Fig.~\ref{fig:pdf_vs_kappa_beta90}).
For isotropic scattering ($\kappa=0$), the Doppler \gls{pdf} is seen to be uniform, regardless of the motion direction.
The exponential trend of the Doppler \gls{pdf} for $\beta=0^\circ$ is evident from Fig.~\ref{fig:pdf_vs_kappa_beta0}, where more concentrated scattering (higher $\kappa$) yields a faster transition and a higher peak value at $f=f_m$, i.e., finite and given by \eqref{eq:dopp_pdf_max}.

As seen from Fig.~\ref{fig:pdf_vs_kappa_beta90}, the behavior of the Doppler \gls{pdf} is different for $\beta=90^\circ$.
As scattering becomes more concentrated in this case, the shape of the Doppler \gls{pdf} becomes similar to that of a zero-mean Gaussian distribution with increasingly smaller variance.
This behavior is determined by the Bessel function in \eqref{eq:dopp_pdf}, as the exponential term is equal to one in this case.
It should be pointed out that the Gaussian shape of the Doppler spectrum is reported for aeronautical channels, e.g., see \cite[Ch.~3.3]{Patzold2012book} and references therein. \looseness=-1

Figs.~\ref{fig:pdf_vs_beta} and \ref{fig:pdf_vs_kappa} also include empirical results obtained through Monte Carlo simulations.
These results are generated by randomly sampling \glspl{doa} from the \gls{vmf} distribution for $10^5$ scatterers, calculating the corresponding Doppler frequencies according to \eqref{eq:dopp_freq}, constructing a histogram with 20 evenly spaced bins and properly normalizing it to yield the empirical Doppler \gls{pdf}.
As we observe, the empirical results align perfectly with the theoretical predictions in all cases, thereby confirming the correctness of the results derived in Section~\ref{sec:dopp_pdf_deriv}.

\section{Conclusions}
\label{sec:conclusions}
The \gls{vmf} distribution is a widely adopted scattering model, particularly in multi-antenna systems performance studies, where accurate modeling of the spatial scattering distribution is crucial.
In this letter we present a simple closed-form expression for the Doppler power spectrum in \gls{vmf} scattering channels, a result not previously available in the literature.

This expression is derived by establishing the relationship between the directional distribution of scatterers on the unit sphere and the Doppler frequency \gls{pdf}, by considering the geometry of the cone encompassing \glspl{doa} associated with the same Doppler frequency shift.
The presented expression accommodates arbitrary mobile velocities, scattering directions and degrees of concentration, with isotropic and deterministic single-point scattering included as special cases.

The obtained result is employed to investigate the impact of the scattering parameters.
The Doppler spectrum is observed to exhibit a Gaussian shape for mobile antenna motion perpendicular to the mean scattering direction, and an exponential one for the parallel motion.



\end{document}

%% file: acronyms.tex
\newacronym[\glslongpluralkey={Directions of Departures}]{dod}{DoD}{Direction of Departure}
\newacronym[\glslongpluralkey={Directions of Arrivals}]{doa}{DoA}{Direction of Arrival}
\newacronym[\glslongpluralkey={Angles of Departure}]{aod}{AoD}{Angle of Departure}
\newacronym[\glslongpluralkey={Angles of Arrival}]{aoa}{AoA}{Angle of Arrival}
\newacronym[\glslongpluralkey={Elevation Angles of Departures}]{eaod}{EAoD}{Elevation Angle of Departure}
\newacronym[\glslongpluralkey={Elevation Angles of Arrivals}]{eaoa}{EAoA}{Elevation Angle of Arrival}
\newacronym[\glslongpluralkey={Azimuth Angles of Departures}]{aaod}{AAoD}{Azimuth Angle of Departure}
\newacronym[\glslongpluralkey={Azimuth Angles of Arrivals}]{aaoa}{AAoA}{Azimuth Angle of Arrival}
\newacronym{mpc}{MPC}{Multipath Component}
\newacronym[first=transmitter (Tx)]{tx}{Tx}{Transmitter}
\newacronym[first=receiver (Rx)]{rx}{Rx}{Receiver}
\newacronym{vmf}{vMF}{von Mises-Fisher}
\newacronym{fb5}{FB5}{Fisher-Bingham}
\newacronym{pdf}{PDF}{Probability Density Function}
\newacronym{cdf}{CDF}{Cumulative Distribution Function}
\newacronym{scf}{SCF}{Spatial Correlation Function}
\newacronym{acf}{ACF}{Auto-Correlation Function}
\newacronym{ccf}{CCF}{Cross-Correlation Function}
\newacronym{psd}{PSD}{Power Spectrum Density}
\newacronym{mimo}{MIMO}{Multiple-Input Multiple-Output}
\newacronym{icas}{ICAS}{Integrated Communication and Sensing}
\newacronym{iid}{i.i.d.}{independent identically distributed}